\documentclass[aps,preprint,prl]{revtex4-1}

\usepackage[utf8]{inputenc}
\usepackage{float}
\usepackage[fleqn]{amsmath}
\usepackage[title]{appendix}
\usepackage{amsmath,geometry, amssymb,verbatim, graphicx, caption,makecell,subcaption,tikz,color,hyperref}
\usepackage{enumitem}
\usepackage{siunitx}
\begin{document}

\title{Slaved Coarsening in Ferronematics}

\author{Aditya Vats}
\author{Varsha Banerjee}
\affiliation{Department of Physics, Indian Institute of Technology Delhi, New Delhi - 110016, India}
\author{Sanjay Puri}
\affiliation{School of Physical Sciences, Jawaharlal Nehru University, New Delhi - 110067, India}

\begin{abstract}
We focus on understanding the influence of the two-component coupling in {\it ferronematics}, a colloidal suspension of magnetic nanoparticles in nematic liquid crystals. Using coarse-grained Landau-de Gennes free energies, we study the ordering dynamics of this complex fluid and present a range of analytical and numerical results. Our main observations are: (i) slaved coarsening for quench temperatures $T$ intermediate to the critical temperatures of the uncoupled components, (ii) slower growth similar to the Lifshitz-Slyozov law ($L \sim t^{1/3}$) for symmetric magneto-nematic coupling, (iii) sub-domain morphologies dominated by interfacial defects for asymmetric coupling strengths. These novel results will serve to guide future experiments on this technologically important system.
\end{abstract}

\maketitle

Liquid crystals (LCs) are a state of matter that is intermediate between conventional solids and liquids with a unique combination of order and fluidity. Nematic liquid crystals (NLCs) are the simplest type, and have a natural tendency to align parallel to one another. This preferred direction introduces strong anisotropy and is described by the nematic director ${\bf n}$. 
The possibility of controlling their optical response by fast reorientation of ${\bf n}$ in a few milliseconds on the application of small electric fields ($1-2$ mV) is the basis of their utility in modern LC displays and optical imaging \cite{Chen_2018}.  However, due to low values of magnetic susceptibility $\sim10^{-6}$ (SI units), large magnetic fields $\sim300$ mT are required to actuate them \cite{DGen_1969}. As a result, most LC devices are mainly driven by electric fields, limiting their applicability in magnetic devices. 

The natural question next is whether the introduction of a small quantity of magnetic material can enhance their sensitivity, thereby introducing the possibility of magneto-optic response in addition to the conventional electro-optic response. This idea was first introduced in 1970 by Brochard and de Gennes in their pioneering work \cite{Broc_1970}, which suggested that the nematic molecules could impose ferromagnetic order in the functionalized magnetic nanoparticles (MNPs) due to surface anchoring even in the absence of an external magnetic field! Although intense experimental efforts were made to create stable ferromagnetic suspensions, it was only in 2013 that Mertelj {\it et al.} obtained the first such using barium hexaferrite magnetic nanoplatelets in pentylcyano-biphenyl LCs  \cite{Mert_Na2013}. Ever since, this fascinating class of materials called {\it ferronematics} (FNs), is enjoying increasing interest from academia and industry as well \cite{ Mert_Soft2014, Qliu_PNAS2016,TPostisk_PRL2017,Mert_PRE2018,Zhng_PRL2015,JSBTai_PNAS2018, Rupn_LC2015, Mert_Ncomm2016,Acke_Nmat2017}. There have been proposals to utilitize the magneto-mechanical and magneto-optic effects for applications in photonics \cite{Acke_Nmat2017}, optical switches \cite{Qliu_PNAS2016}, complex fluids \cite{TPostisk_PRL2017,Mert_PRE2018}, and even particle physics and cosmology \cite{JSBTai_PNAS2018}.

The primary quest after the creation of stable FNs has been to understand the consequences of coupling between the nematic and magnetic components on their equilibrium and non-equilibrium properties. There have been some intriguing experimental observations in this direction. For instance, Shuai et al. could use the magneto-nematic coupling to spontaneously create flux closure loops which were sensitive to even the Earth's magnetic field \cite{Mert_Ncomm2016}. In another contribution, Mertelj and Lisjak cooled a ferronematic drop cooled from an isotropic phase to the nematic phase and observed domain growth in the presence of a filed appleid along {\bf n}  \cite{Mert_LCR2017}. Using magneto-optic techniques, they detected several {\it bubbles} or {\it small domain morphologies} with magnetization parallel or anti-parallel to the field which coarsened with time. The domain growth was always accompanied by the flow of the suspension and motion of the defect lines.  Mertelj and Lisjak thus demonstrated that the mechanism of domain growth was mediated via a magneto-nematic interaction \cite{Mert_PRE2018}.

The problem of coarsening (or domain growth) after a quench from a disordered phase ($T>T_c$) to an ordered phase ($T<T_c$) holds a special appeal in non-equilibrium physics \cite{Bray_2002,Puri_2009}. If the morphology of the coarsening domains is unchanged in time, the system exhibits dynamical scaling and is characterized by a unique divergent length scale $L(t)$. The growth law reveals important details of the free-energy landscape and the relaxation (response) time-scales in the system. In pure and isotropic systems, $L(t) \sim t^{1/z}$, where the growth exponent $1/z$ depends on various factors such as conservation laws, defects and flow fields. For example, systems with non-conserved kinetics obey the Lifshitz-Allen-Cahn (LAC) law: $L(t) \sim t^{1/2}$ which is characteristic of systems with no energy barriers to coarsening \cite{Allen_Cahn_1979}. Systems with disorder and competing interactions have a complicated free-energy landscape and a plethora of relaxation time-scales. Domain growth in these systems exhibits logarithmic behavior in the asymptotic limit \cite{Puri_Zannetti_2010,Puri_Zannetti_2012}. What insights can a coarsening experiment provide for FNs, which are described by {\it two} coupled order parameters? What is the influence of coupling strengths on growth laws? What happens if only one of the two components is in an ordered phase? Motivated by these and related questions, we develop a time-dependent Ginzburg-Landau (TDGL) formulation using coarse-grained free energies to study non-equilibrium properties of FNs rendered unstable after a thermal quench \cite{Bray_2002,Puri_2009}. 
This letter focuses on understanding the influence of the two-component coupling. We present a range of analytical and numerical results for this problem. This benchmarking study will serve to guide future experiments on this technologically important system. 

We report three novel results which are as follows:
(i) For {\it shallow} quenches ($T_c^N<T<T_c^M$, where $N$ and $M$ refer to nematic and magnetic), the ordering magnetic component {\it enslaves} the nematic component to coarsen. Their domains are co-aligned. Similar statements hold for quenches such that $T_c^M<T<T_c^N$.  
(ii) Depending on the nature of coupling, domain growth can obey the Lifshitz-Allen-Cahn law $L(t)\sim t^{1/3}$, or a slower $L(t)\sim t^{1/3}$ growth usually referred to as the Lifshitz-Slyozov  law. The latter is surprising as it usually characterizes conserved kinetics \cite{Bray_2002,Puri_2009}, whereas both order parameters here are non-conserved. 
(iii) The structure factor $S(k)$ for the domain morphologies exhibits Porod decay, $S(k)\sim k^{-(d+1)}$, which is characteristic of scattering from sharp interfaces. This contradicts our naive expectation of the generalized Porod tail $S(k)\sim k^{-(d+2)}$ for scattering from vortex defects in continuous-spin models \cite{Bray_1992}.

In the Landau-de Gennes (LdG) free energy description, the FN is described by two order parameters: (i) the ${\bf Q}$-tensor, which contains information about the orientational order of the LC \cite{Mtram_ArX_2014}; and (ii) the magnetization vector ${\bf M}$, which is the magnetic moment of the suspended nanoparticles. We will focus on the 2-dimensional case here, which is relevant for applications such as LC displays for instance. We allow ${\bf M}$ to have a variable magnitude, including ${\bf M} =0$ to capture segregation effects. In $d=2$, the ${\bf Q}$-tensor is a $2\times 2$ matrix with elements $Q_{ij}=S(\hat{n_i}\hat{n_j}-\delta_{ij}/2)$. The scalar order parameter $S$ measures the fluctuations about the leading eigenvector ${\bf n}$ \cite{AMaj_QT2012}. Further, $\mbox{ Tr} {\bf Q} = 0$, $\mbox{ Tr} {\bf Q}^2 = 2(Q_{11}^2+Q_{12}^2) = S^2/2$ and  $\mbox{ Tr} {\bf Q}^3 = 0$. The free energy for the ferronematic system has been modelled as \cite{JP_DG_1995,HPlein_2001}:
\begin{eqnarray}
\label{F_e}
G(\mathbf{[Q,M]}) &=& \int \mbox{d}\bf{r} \bigg[ \frac{A}{2}\mbox{Tr}(\boldsymbol{Q}^2)+\frac{B}{4}\mbox{Tr}(\boldsymbol{Q}^4)+\frac{L}{2} |\nabla{\bf Q}|^2 \nonumber\\ 
&& +\frac{\alpha}{2}\left|\boldsymbol{M}\right|^2+\frac{\beta}{4}\left|\boldsymbol{M}\right|^4 + \frac{\kappa}{2}\left|\nabla \boldsymbol{M}\right|^2 - \frac{\gamma\mu_0}{2} \sum_{i,j=1}^{2}Q_{ij}M_iM_j\bigg].
\end{eqnarray}
The first three terms represent the LdG free energy for the nematic component , the next three correspond to the Ginzburg-Landau free energy for the magnetic component and the last term represents the magneto-nematic coupling. To leading order, the coupling term is taken to be the dyadic product of {\bf Q}-tensor and {\bf M} to respect the rotational invariance of the free energy. 
The Landau coefficients $ A = A_0(T-T_{c}^N)$ and $ \alpha = \alpha_{0}(T-T_{c}^M)$, where $A_0$ and $\alpha_0$ are positive constants. The parameters $B$ and $\beta$ are positive material-dependent constants, $L$ and $\kappa$ are the elastic constants, and $\gamma$ and $\mu_0$ are the coupling strength and the magnetic permeability respectively. These phenomenological parameters can be estimated from experimentally measured quantities. 

The dissipative dynamics of the ferronematic is studied using the coupled time-dependent Ginzburg-Landau (TDGL) equations: $ \partial \psi/\partial t = -\Gamma_{\psi}\delta G[\mathbf{Q,M}]/\delta \psi$, where the terms on the right are the functional derivatives of the free energy functional $G[\mathbf{Q,M}]=\int \mbox{d}\textbf{r} g(\mathbf{Q,M})$ \cite{Puri_2009}.  A dimensionless form of the TDGL equations can be obtained by introducing rescaled variables $\mathbf{Q} = c_N\mathbf{Q^\prime}$, $\mathbf{M} = c_M\mathbf{M^\prime}$, $\textbf{r} = c_r\textbf{r}^\prime$,  $t=c_tt^\prime$. Dropping the primes yields the following equations:
\begin{eqnarray}
\label{TDGL1}
    \frac{1}{\Gamma}\frac{\partial Q_{11}}{\partial t}&=&\pm Q_{11}-(Q_{11}^2+Q_{12}^2)Q_{11}+l\nabla^2Q_{11}+c_1[M_1^2-M_2^2],\\
\label{TDGL2}
    \frac{1}{\Gamma}\frac{\partial Q_{12}}{\partial t}&=&\pm Q_{12}-(Q_{11}^2+Q_{12}^2)Q_{12}+l\nabla^2Q_{12}+2c_1[M_1M_2],\\
\label{TDGL3}
    \frac{\partial M_1}{\partial t}&=&\pm M_1-|\mathbf{M}|^2M_1+\nabla^2M_1+c_2[Q_{11}M_1+Q_{12}M_2],\\
\label{TDGL4}
    \frac{\partial M_2}{\partial t}&=&\pm M_2-|\mathbf{M}|^2M_2+\nabla^2M_2+c_2[Q_{12}M_1-Q_{11}M_2].
\end{eqnarray}
The dimensionless parameters in Eqs.~(\ref{TDGL1})-(\ref{TDGL4})  are
\begin{equation}
\label{constants}
    c_1=\frac{\gamma \mu_0 |\alpha|}{4|A|\beta}\sqrt{\frac{2B}{|A|}},
   \\
 c_2=\frac{\gamma \mu_0 }{|\alpha|}\sqrt{\frac{|A|}{2B}},
\\
     l=\frac{|\alpha|L}{2|A|\kappa},
\\
    \Gamma= \frac{2|A|\Gamma_Q}{|\alpha|\Gamma_M}.
\end{equation}
The $\pm$ sign with the first terms on the right depend on whether the corresponding component is above ($-$) or below ($+$) its critical temperature. The parameters $c_1$ and $c_2$ are rescaled coupling constants, $l$ sets the scale for relative diffusion of the nematic and magnetic components, and  $\Gamma$ is the relative damping coefficient. For simplicity, we set $l=1$  and $\Gamma=1$. Therefore, the only parameters in our model are $c_1$ and $c_2$. We emphasize that these originate from the same coupling term in Eq.~(\ref{F_e}). However in our dimensionless rescaling, they are combined with factors which determine the dimensional scales of the order parameters $\textbf{Q}$ and $\textbf{M}$ (see Eq.~(\ref{constants})).

There are three interesting cases in this problem: (1) $T_c^M>T>T_c^N$, (2) $T_c^N>T>T_c^M$, and  (3) $T<\mbox{min}\{T_c^N,T_c^M\}$. We study these for the following sub-cases below: (i) $c_1\ne0$, $c_2=0$, (ii) $c_1 = 0$, $c_2\ne0$, and (iii) $c_1=c_2=c$, as specified in Table I. A few remarks about the limiting cases are in order. Asymmetric coupling is not unusual, as the order parameters can have vastly different magnitudes in experiment, e.g., large magnetic particle in a bath of small LC molecules. We do not expect to precisely realize $c_1$ or $c_2=0$, or $c_1=c_2=c$ in experiments. However, due to their tractability, the limiting cases (i)-(iii) provide useful guidelines for experimental studies. We have numerically solved Eqs.~(\ref{TDGL1})-(\ref{TDGL4}) by implementing an isotropic Euler discretization on an $N^2$ lattice ($N = 1024$) with periodic boundary conditions in both directions \cite{kin_Num2009}. The discretization mesh sizes were $\Delta t = 0.01$ and $\Delta x = 1.0$. We present here results for prototypical cases of Table I, and provide a detailed analysis for all the cases in the supplementary material.

Although our primary interest is in coarsening, we first determine the stationary solutions $({\bf Q^*,M^*})$ by setting the time and space derivatives to zero in Eqs.~(\ref{TDGL1})-(\ref{TDGL4}). We then perform a linear stability analysis  by studying the growth of initially small fluctuations ${\textbf Q}^*(\textbf{r},0) = {\textbf Q}^*+\delta{\textbf Q}^*(\textbf{r},0)$; ${\textbf M^*}(\textbf{r},0) = {\textbf M^*}+\delta {\textbf M^*}(\textbf{r},0)$. We provide the stable solutions for  all the cases in Tables S1 - S3 in the supplementary material. They serve as the framework to interpret the non-equilibrium evolution of the FN after a temperature quench. Fig.~1 shows the morphologies of the nematic (left) and magnetic (right) components for Case 1 (i) of Table I. The snapshots are shown at $t=10^3$ with $c_1=4$. In the nematic picture, blue (black)  corresponds to {\bf n} in the first (or third) quadrant,  while green (light gray) corresponds to {\bf n} in the second (or fourth) quadrant. In the magnetisation picture, the colours blue (black), red (dark gray), green (light gray) and yellow (white) denote $\textbf{M}$ lying in the first, second, third and fourth quadrant respectively. We stress that $\textbf{M}$ is always parallel {\bf n}. The magnetic domains coarsen as expected, but what is unusual is the {\it slaved coarsening} of the nematic phase, and its co-alignment with {\bf M}. We also find that the magnitudes of {\bf Q} and {\bf M} are in accordance with the analytical values for the corresponding stable stationary solution in Table S1. 

To quantify the morphologies and domain growth, we define the characteristic length scale $L(t)$ as the distance over which  the correlation function decays to (say) 0.2 times its maximum value. If the ordering system is isotropic and characterized by a single length scale, then the correlation function obeys dynamical scaling: $C({\textbf{r}},t) = f\left(r/L\right)$, where $f(x)$ is a scaling function.  An equivalent probe is the structure factor $S(\textbf{k},t)$, which is the Fourier transform of $C({\textbf{r}},t)$, and is usually obtained in small-angle scattering experiments. The corresponding dynamical-scaling form is $S(\textbf{k},t) = L^{d}\tilde{f}\left(kL\right)$, where $\tilde{f}(p)$ is the Fourier transform of $f(x)$ \cite{Bray_2002,Puri_2009}. Our two-component system has two length scales $L_{Q}$ and $L_M$, characterizing domain growth of the nematic and magnetic components, respectively. For pure nematic and magnetic systems ($c_1=c_2=0$), it is well established that the components obey the LAC law: $L(t)\sim t^{1/2}$ in $d>2$ and  $L(t)\sim \left[t/\ln t\right]^{1/2}$ in $d=2$. 

In Fig.~2, we show the growth laws for Case 1(\romannumeral 1) with $c_1 =3, \ 4, \ 5$ and $c_2 = 0$ (left frame); and Case 1(\romannumeral 3) with $c_1=c_2=c$ for $c=3, \ 4, \ 5$ (right frame). The solid symbols denote $L_{Q}(t)$ vs. $t$ while the open symbols denote $L_M(t)$ vs. $t$. The linear variation on the log-log scale suggests power laws: $L(t)\sim t^{1/z}$. For accurate determination of the slopes, we evaluated the {\it effective} growth exponent $\bar{z} =\partial \ln t/\partial \ln L(t)$. The insets in Figs.~2(a)-(b) show $\bar{z}$ vs. $t$ on a semi-log scale. The dashed lines indicate the exponent values $z=2$ and $3$. The data for Case 1(\romannumeral 3) is consistent with the LAC law for both components. Thus the slaved nematic order parameter, which is naturally isotropic, is driven to ordering by the magnetisation field. Both these fields show the same growth law. We emphasise that the exponent $z\gtrsim 2$ because of the logarithmic correction arising in the growth law for $d=2$.

The data for Case 1(\romannumeral 3) in Fig. 2(b) is the second unexpected outcome of our study. Both the systems indeed coarsen together, but now the growth law is much closer to $L(t)\sim t^{1/3}$. The latter, usually referred to as the Lifshitz-Slyozov (LS) law, is characteristic of systems with conserved dynamics though both order parameters are non-conserved in the present case! Our study indicates that the LS-like law is observed in all three cases for a symmetric coupling between components, see Tables S1-S3. 

The nature of the defects can be interpreted from the tail of the structure factor \cite{Bray_Puri_1991}. For the cases Case 1(\romannumeral 1) and Case 1(\romannumeral 3), we find that the nematic and magnetic components  exhibit the {\it generalized Porod decay} $S(k,t) \sim k^{-(d+n)} = k^{-4}$ for $d=2$, $n=2$. This is characteristic of scattering from vortex defects in $XY$-type spin models \cite{Bray_2002}. As a matter of fact, we always observe the $k^{-4}$ decay for (i) $c_1\ne0$, $c_2= 0$, and (iii) $c_1=c_2=c$, see Tables S1-S3.

Finally, we present some results for the situation when the nematic field evolves freely but the magnetisation field is driven by the nematic field, i.e., $c_1= 0$, $c_2\ne 0$. We focus on Case 3(\romannumeral 2) of Table I with $T<\mbox{min}\{T_c^N,T_c^M\}$ so that both the components are in the ordered phase. The corresponding  results are provided in Fig.~3.  In (a) and (b) we show the nematic and magnetic morphologies for $c_2=4$ and $t=10^3$. We see that the components are aligned, i.e., ${\bf n} \parallel {\bf M}$. However, the magnetic component exhibits a small sub-domain morphology (SDM) due to the two possible alignments, ${\bf n}\parallel {\bf M}$ and ${\bf n}\parallel -{\bf M}$, with the same energy. The sub-domain gradients in ${\textbf M}$ have a cost in terms of the surface tension, but this is estimated to be negligible compared to the entropic gain due to the formation of the SDM. The magnitudes of {\bf Q} and {\bf M} agree with the stable stationary solutions provided in Table S3. In Fig.~3(c), we depict the growth laws for $c_2=3, \ 4, \ 5$. While $L_{Q}(t)\sim t^{1/2}$ as expected,  $L_M(t)$ saturates to $L_M^S$ due to the formation of the SDM. In the uncoupled limit $c_2\rightarrow 0$, we expect $L_M^S \rightarrow \infty$. More insights on the SDM are provided by the scaled structure factor, $L^{-2}S(k,t)$ vs. $kL$, plotted in Fig.~3(d) for $c_2=4$. As expected, $S_{Q}(k,t)$ exhibits a generalized Porod tail $S_Q\sim k^{_4}$, due to scattering off vortex-like defects in Fig.~3(a). However, $S_M(k,t)$ shows the usual Porod tail $S_M\sim k^{-3}$! This is a result of scattering from the sharp ``interfaces'' between the sub-domains with magnetization ${\textbf M}$ and $-{\textbf M}$. Though $\textbf{M}$ is a continuous order parameter, the nematic coupling enforces a discrete up-down symmetry for ${\textbf M}$ in the SDM. We find that the SDM and the $k^{-3}$ law exhibited by the magnetic component is generic to $c_1= 0$, $c_2\ne0$, see Tables S1-S3.

So what are the novel insights from this first coarsening study of a ferronematic? Our TDGL formulation for the FN has allowed us to understand the effects of magneto-nematic coupling on morphologies and growth laws. Rather than the nature of the quench, e.g., shallow (say $T_c^N<T<T_c^M$) or deep $\left(T<\mbox{min}\{T_c^N,T_c^M\}\right)$, it is the relative coupling strengths $c_1$ which dictate the systemic behavior. There are three novel observations from our study: (i)  slaved coarsening for quench temperatures $T$ between the critical temperatures of the uncoupled components, (ii) slower growth similar to Lifshitz-Slyozov $\left(L(t) \sim t^{1/3}\right)$ law for symmetric magneto-nematic coupling ($c_1=c_2=c$), (iii) sub-domain morphologies dominated by interfacial defects for asymmetric coupling strengths ($c_1=0, c_2\ne0$).  

Finally, what is the experimental relevance of this simplistic model? The Landau coefficients $A$, $B$ and $L$ are related to experimentally measured quantities like the critical temperature, the latent heat of transition and the order parameter \cite{Priestly_2012}. Similarly, coefficients $\alpha$, $\beta$ and $\kappa$ can be evaluated from the measurements of magnetization and susceptibility \cite{Hberg_2015}. The coupling constant $\gamma$, in the experiments of Mertelj et al., has been estimated from the reversal fields of hysteresis loops \cite{Mert_Na2013}. It is therefore possible to experimentally determine our dimensional scales and the dimensionless coupling constants $c_1$ and $c_2$. We hope our theoretical results will propose and guide coarsening experiments in FNs. This is an emergent field of research, and combined experimental and theoretical efforts are needed  to understand this fundamentally rich and technologically important system. Our work is a step in this direction.

\begin{table}
\begin{center}
      \begin{tabular}{|c|c|}
       \hline
       Quench temperature & Coupling limits
        \\ 
       \hline
        \makecell{1) $T_c^N<T<T_c^M$ \\ 2) $T_c^M<T<T_c^N$ \\ 3) $T<\mbox{min}\{T_c^N,T_c^M\}$} & \makecell{(\romannumeral 1) $c_1\ne0$,\ $ c_2=0$ \\(\romannumeral 2)\ $c_1 = 0$,\ $ c_2\ne0$\\(\romannumeral 3) $c_1=c_2=c$\hspace*{0.3cm} } \\
        \hline
    \end{tabular}
    \end{center}
    \caption{Coarsening studies which we have undertaken.}
    \label{Tab1}
\end{table}

Supplementary material contains Tables S1, S2, S3 summarizing stable solutions, growth laws and structure factor tail behavior for Cases 1, 2, 3 in the different coupling limits described in Table I.

AV acknowledges UGC, India for a research fellowship. AV and VB gratefully acknowledge partial financial support from DST-UKIERI and the HPC facility of IIT Delhi for the computational resources.

\bibliographystyle{h-physrev}
\bibliography{ref}

\newpage

\begin{figure}[H]
                            \centering                      
                            \includegraphics[width=0.75\linewidth]{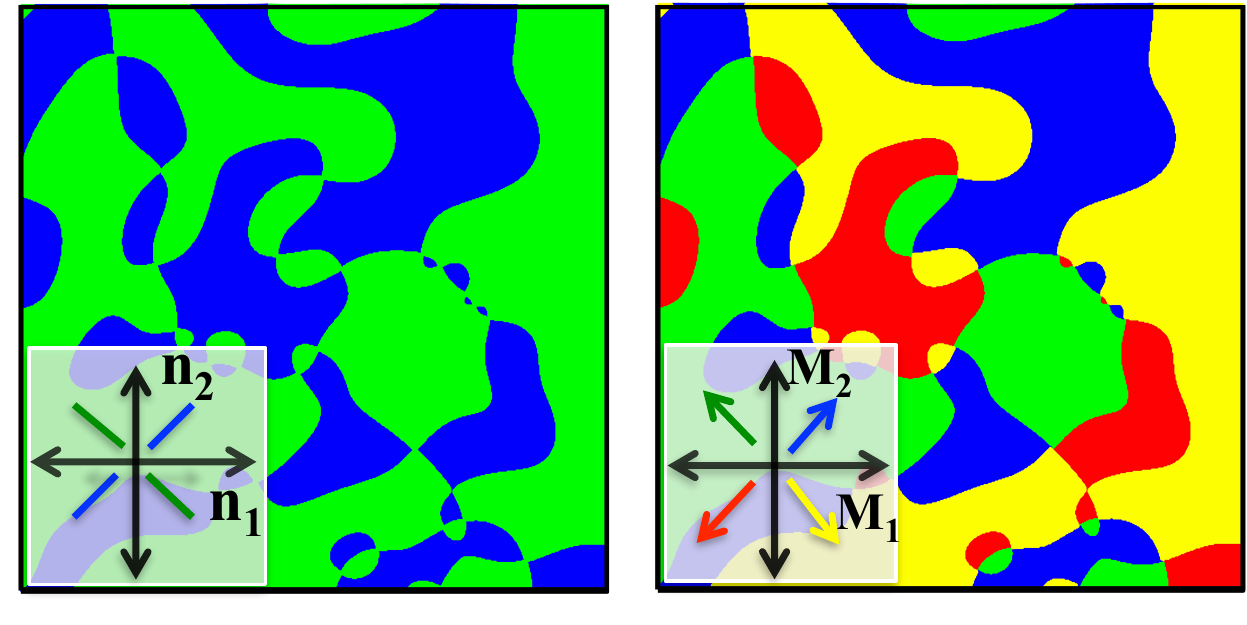} 
                          \caption{Morphology snapshots corresponding to Case 1(\romannumeral 1) of Table I for the nematic (left) and magnetic (right) components at time $t=10^3$ with coupling constants $c_1=4,\ c_2=0$. The colour code shown in the insets is detailed in the text.}
                           \label{mp_case1}
                   \end{figure}

\begin{figure}[H]
                            \centering                      
                            \includegraphics[width=0.75\linewidth]{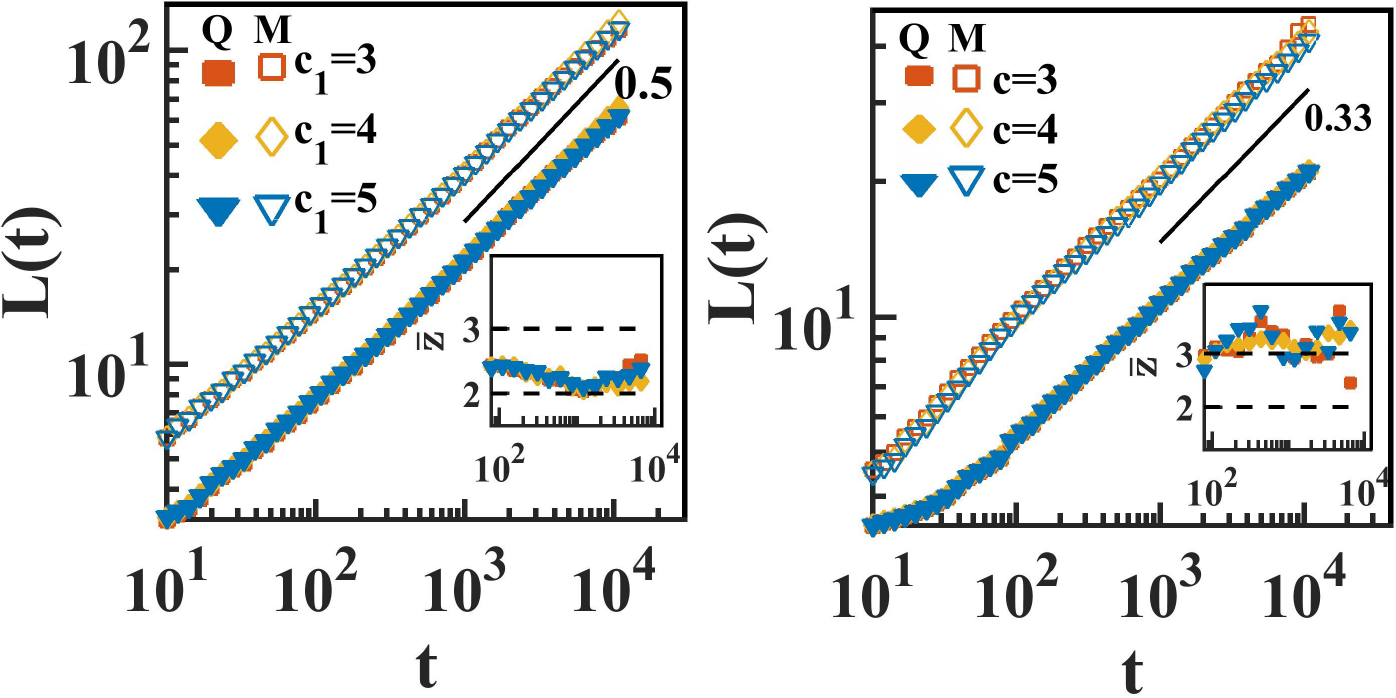} 
                             \caption{Growth laws on a log-log scale for Case 1(\romannumeral 1) (left) and Case 1(\romannumeral 3) (right) of Table I. The solid (open) symbols denote the nematic (magnetic) component. The insets show the effective growth exponent $\bar{z}=[d(\ln L)/d(\ln t)]^{-1}$ vs. $t$. The dashed lines corresponds to $\bar{z}=2$ (left) and $\bar{z}=3$ (right).}
                           \label{gl_case1}
                   \end{figure}

\begin{figure}[H]
                            \centering                      
                            \includegraphics[width=0.75\linewidth]{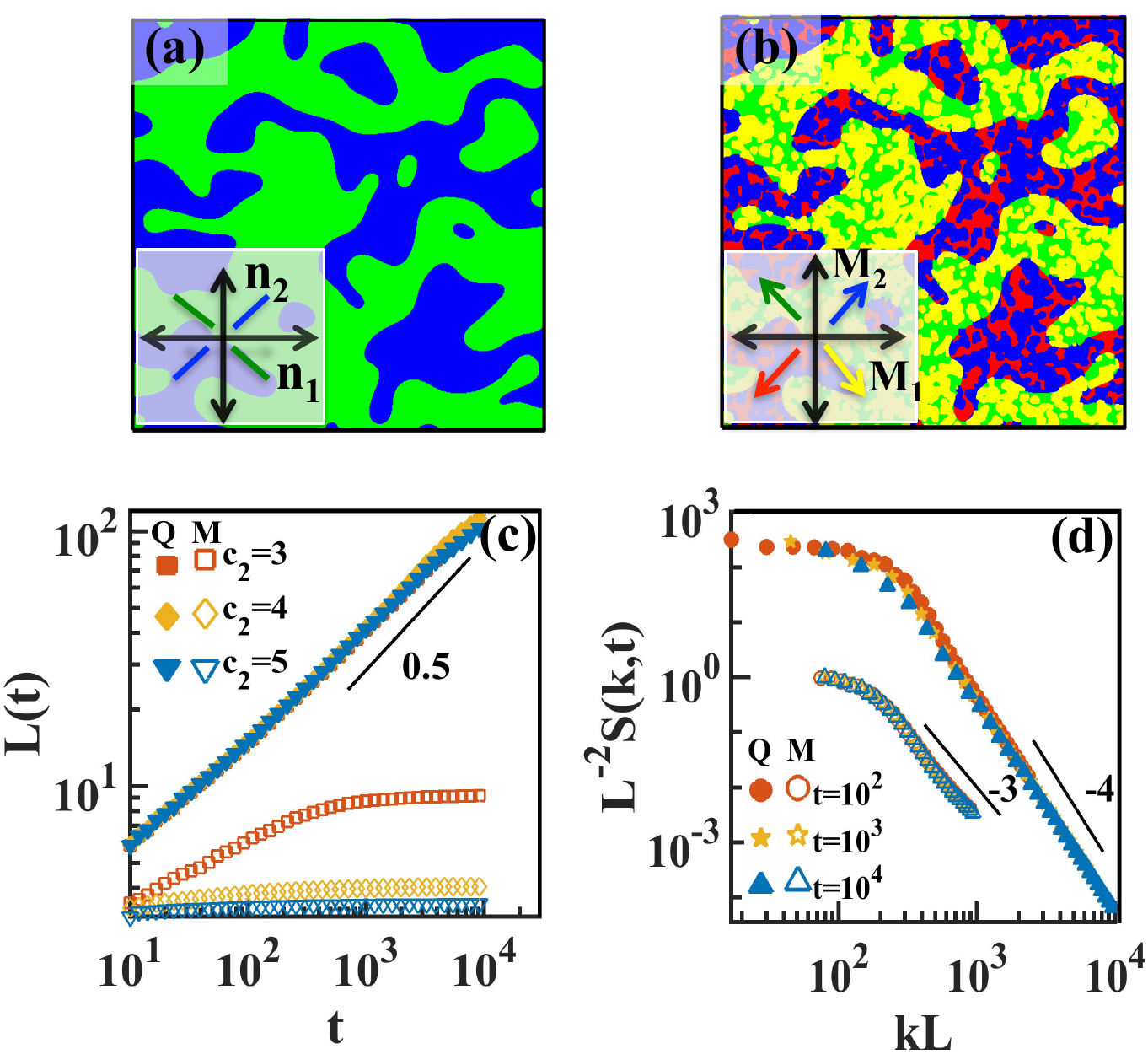} 
                           \caption{(a) Nematic and (b) magnetic morphologies  for the Case 3(\romannumeral 2) of Table I with $c_1=0,\ c_2=4$. Growth laws $L(t)$ vs. $t$ for specified values of $c_2$ are shown in (c). The scaled structure factor data, $L^{-2}S(k,t)$  vs. $kL$, at the specified times $t$ is shown in (d). The solid (open) symbols in (c) and (d) denote the nematic (magnetic) component.}
                           \label{mp_gl_case3}
                   \end{figure}
\newpage
\setcounter{table}{0}
\renewcommand{\thetable}{S\arabic{table}}
\section*{Supplementary Material}
\subsection*{TABLES}
 \begin{center}   
\begin{table}[h]
     \centering
      \begin{tabular}{|c|c|c|c|}
            \hline
                \textbf{Coupling limits}& \makecell{ \textbf{Stable stationary solutions:}\\($M_1^*,M_2^*,Q_{11}^*,Q_{12}^*$)}&\makecell{\textbf{Growth laws}\\ ($L_M,L_Q)$}&\makecell{\textbf{Structure Factor Tails}\\ ($S_M,S_Q$)}\\ 
            \hline
   
                (\romannumeral 1)\ $c_1\ne0$,\ $ c_2=0$ &  \makecell{($1,0,r_Q,0$) \\ $r_Q=c_1(1+\hat{S})^{-1}$\\ $\hat{S}=S^2/4={3}^{-1}(-2+a_1+{a_1}^{-1})$\\ $a_1={2^{1/3}}(a_2+c_1\sqrt{54+27a_2})^{-1/3}$\\ $a_2=2+c_1^2$\\ } & $(t^{1/2},t^{1/2})$  &$(k^{-4},k^{-4})$\\
           \hline
               (\romannumeral 2)\ $c_1= 0$,\ $ c_2\ne0$ & $(1,0,0,0)$& $(t^{1/2},\mbox{No growth})$  &$(k^{-4},\mbox{Uniform})$\\
           \hline
              (\romannumeral 3)\ $c_1=c_2=c$\hspace*{.2cm}&\makecell{($r_M,0,r_Q,0$)\\
              $r_m=\big[(1+\hat{S})(1+c^2+\hat{S})^{-1}]^{1/2}$\\
              $r_Q= c\ (1+\hat{S}+ c^2)^{-1}$\\
              $\hat{S}=S^2/4=(1+c^2)(a_1-2/3)+{a_1}^{-1}$\\
              $3a_1={2^{1/3}}{(33c^2+a_2+{a_3}^{1/2})^{-1/3}}$\\
              $a_3=1053c^4+54c^2a_2$\\
              $a_2=2c^2+6c^4+2c^6$} & $(\sim t^{0.33},\sim t^{0.33})$& $(k^{-4},k^{-4})$ \\
       \hline
       
         \end{tabular}
     \caption{Coupling limits, stable stationary solutions, growth laws and structure factor tails for Case 1 of Table I corresponding to a quench at temperature $T$ such that $T_c^{N}<T<T_c^{M}$.}
 
    \label{TabS1}
\end{table}
\end{center}

\begin{center}
    
\begin{table}[h]
    \centering
      \begin{tabular}{|c|c|c|c|}
           \hline
                \textbf{Coupling limits}& \makecell{ \textbf{Stable stationary solutions:}\\($M_1^*,M_2^*,Q_{11}^*,Q_{12}^*$)}&\makecell{\textbf{Growth laws}\\ ($L_M,L_Q$)}&\makecell{\textbf{Structure Factor Tails}\\ ($S_M,S_Q$)}\\ 
            \hline
                (\romannumeral 1)\ $c_1\ne0$,\ $ c_2=0$ & $(0,0,1,0)$& $(\mbox{No growth},t^{1/2})$  &$(\mbox{Uniform},k^{-4})$\\
           \hline
                 ({\romannumeral 2})\ $c_1=0$,\ $ c_2\ne0$ &  \makecell{($r_M,0,1,0$) \\ $r_m=\sqrt{c_2-1}$\\} &  $(\mbox{Saturation},t^{1/2})$  & $(k^{-3},k^{-4})$\\
           \hline
              (\romannumeral 3)\ $c_1=c_2=c$\hspace*{.2cm}&\makecell{($r_M,0,r_Q,0$)\\
              $r_m=\big[(\hat{S}-1)(1+c^2-\hat{S})^{-1}]^{1/2}$\\
               $r_Q= c\ (1-\hat{S} + c^2)^{-1}$\\
              $\hat{S}=S^2/4=0.5(2+c^2+\sqrt{4c^2+c^4})$\\} &  $(\sim t^{0.33},\sim t^{0.33})$  & $(k^{-4},k^{-4})$   \\
         \hline
     \end{tabular}
    \caption{Coupling limits, stable stationary solutions, growth laws and structure factor tails for Case 2 of Table I corresponding to a quench at temperature $T$ such that $T_c^{M}<T<T_c^{N}$.}
 \end{table}
 \label{TabS2}
\end{center}
\begin{center}
    
\begin{table}[h]

    \centering
      \begin{tabular}{|c|c|c|c|}
              \hline
                    \textbf{Coupling limits}& \makecell{ \textbf{Stable stationary solutions:}\\($M_1^*,M_2^*,Q_{11}^*,Q_{12}^*$)}&\makecell{\textbf{Growth laws}\\ ($L_M,L_Q$)}&\makecell{\textbf{Structure Factor Tails}\\ ($S_M,S_Q$)}\\ 
              \hline
            
                   (\romannumeral 1)\ $c_1\ne0$,\ $ c_2=0$ &  \makecell{($1,0,r_Q,0$) \\ $r_Q=-c_1(1-\hat{S})^{-1}$\\ $\hat{S}=S^2/4={1}/{3}(-2+a_1+a_1^{-1})$\\ $a_1={2^{1/3}}{(a_2+c_1\sqrt{-54+27a_2})^{-1/3}}$\\} & $(t^{1/2},t^{1/2})$  & $(k^{-4},k^{-4})$ \\
              \hline
     
                  (\romannumeral 2)\ $c_1=0$,\ $ c_2\ne0$ &  \makecell{($r_M,0,1,0$) \\$r_M=\sqrt{c_2+1}$\\} & $(\mbox{Saturation},t^{1/2})$ & $(k^{-3},k^{-4})$ \\
              \hline
                 (\romannumeral 3)\ $c_1=c_2=c$\hspace*{0.2cm}&\makecell{($r_M,0,r_Q,0$)\\ $r_m=\big[(1-\hat{S})(1+c^2-\hat{S})^{-1}]^{1/2}$\\ $r_Q= c\ (1-\hat{S} + c^2)^{-1}$\\ $\hat{S}=S^2/4=0.5(2+c^2+\sqrt{4c^2+c^4})$} & $(\sim t^{0.33},\sim t^{0.33})$ & $(k^{-4},k^{-4})$  \\
         \hline 
         \end{tabular}
    \label{TabS3}
\caption{Coupling limits, stable stationary solutions, growth laws and structure factor tails for Case 3 of Table I corresponding to a quench at temperature $T$ such that $T<\mbox{min}\{T_c^{M},T_c^{N}\}$.}
    \end{table}

\end{center}

\end{document}